\documentclass[%
reprint,
 amsmath,amssymb,
 aps,
pre,
]{revtex4-2}

\usepackage{graphicx}
\usepackage{dcolumn}
\usepackage{bm}
\usepackage{xcolor}
\usepackage{soul}
\newcommand{\fig}[1]{Fig.~\ref{#1}}
\newcommand{\eq}[1]{Eq.~(\ref{#1})}

\newcounter{nameOfYourChoice}

\begin{document}

\title{Viscous liquid dynamics modeled as random walks within overlapping hyperspheres}

\author{Mark F. B. Railton}\email{mfbr@ruc.dk}
\author{Eva Uhre}
\author{Jeppe C. Dyre} 
\author{Thomas B. Schr{\o}der} \email{tbs@ruc.dk}
\affiliation{%
Glass and Time, IMFUFA, Department of Science and Environment, Roskilde University, P.O. Box 260, 4000 Roskilde, Denmark.}

\date{\today}

\begin{abstract}
The hypersphere model is a simple one-parameter model of the potential energy landscape of viscous liquids, which is defined as a percolating system of same-radius hyperspheres randomly distributed in $\mathbb{R}^{3N}$ in which $N$ is the number of particles. We study random walks within overlapping hyperspheres in 12 to 45 dimensions, i.e., above the percolation threshold, utilizing an algorithm for on-the-fly placement of the hyperspheres in conjunction with the kinetic Monte Carlo method. We find behavior typical of viscous liquids; thus decreasing the hypersphere density (corresponding to decreasing the temperature) leads to a slowing down of the dynamics by many orders of magnitude. The shape of the mean-square displacement as a function of time is found to be similar to that of the Kob-Andersen binary Lennard-Jones mixture and the Random Barrier Model, which predicts well the frequency-dependent fluidity of nine glass-forming liquids of different chemistry [Bierwirth \textit{et al.}, Phys. Rev. Lett. \textbf{119}, 248001 (2017)].
\end{abstract}

\maketitle

\section{\label{sec:level1}Introduction}
Experimental studies\cite{OlsenNB2001Tsiv,NielsenAlbenaI.2009Poat, GabrielJan2017DPa,Pabst20213685,SidebottomD.L.2023Gri,GainaruCatalin2019Ssso,BierwirthSPeter2017GPMR} of the dynamics of glass-forming liquids suggest that there might exist a generic $\alpha$-relaxation in viscous liquids sufficiently close to the glass transition. Specifically, one has searched for universalities\cite{hec18} in the linear-response properties probed, e.g., by the frequency-dependent dielectric loss \cite{ric15}, shear or bulk moduli \cite{chr95,hec13,gun14}, or specific heat \cite{nie96,gun11}. One suggested such universality is the conjecture that the high-frequency loss varies with frequency $\omega$ as $\omega^{-1/2}$\cite{OlsenNB2001Tsiv,NielsenAlbenaI.2009Poat,BEL,alma99122800221505763,MontroseC.J.1970SDiL,cun88,MajumbarC.K.1971Srfo,GlarumSivertH.1960DRoI,SJOGRENL1990Tdov}. 

Another more recently suggested possible universality refers to the random barrier model (RBM). Using experimental frequency-dependent shear-modulus data it was demonstrated\cite{BierwirthSPeter2017GPMR} that the real part of the frequency-dependent fluidity (inverse dynamic viscosity) for nine glass-forming liquids of different chemistry is well described by the RBM. 
For a crystallization-resistant version of the Kob-Andersen binary Lennard-Jones mixture, long molecular dynamics simulations demonstrated\cite{SchrderThomasB2020Smdi} that the master curve for the mean-square displacement (MSD) at low temperatures is also well fitted by the RBM prediction.

The RBM was originally proposed for ac conduction in disordered solids modeled as a random walk on a cubic lattice with identical site energies and random energy barriers for nearest-neighbor jumps\cite{DYREJ.C1988Trfb, SchroderTB2008PRL}. Thus the RBM is based on a three-dimensional potential energy landscape that is a lot simpler than the high-dimensional landscapes of viscous liquids. In such liquids, the potential energies of the inherent structures (local potential energy minima, corresponding to the lattice sites of the RBM) are known to have a broad distribution\cite{BuchnerStephan1999Pelo,SAKSAENGWIJITA2004Ootf,StarrFW2001Tasa}. This makes it puzzling that the RBM describes the data so well, since it is characterized by identical site energies. 

In this paper, we study a simple model of the constant-potential-energy hypersurface $\Omega$ defined in configuration space as follows

\begin{equation}
        \Omega=\{ (\textbf{r}_1,...,\textbf{r}_N)\in \mathbb{R}^{3N}\ |\ U(\textbf{r}_1,...,\textbf{r}_N) = U_0\}\,.
\end{equation}
Here $U_0$ denotes the constant potential energy, $N$ the number of particles, and $U(\textbf{r}_1,...,\textbf{r}_N)$ the potential energy as a function of particle coordinates $\textbf{r}_1,...,\textbf{r}_N$. 

Samuelsen \textit{et al.}\cite{https://doi.org/10.48550/arxiv.2206.03000} have proposed a simple one-parameter toy model of $\Omega$, which we will refer to as the ``hypersphere model''. The hypersphere model is defined\cite{https://doi.org/10.48550/arxiv.2206.03000} as the surface of a percolating system of equally sized hyperspheres centered on independently and randomly distributed points in a $d$-dimensional hypercube with periodic boundary conditions. Note that this model has no reference to the physical spatial dimension $d=3$. 

One motivation for studying this model is that Samuelsen \textit{et al.} found dynamics very similar to the RBM and the Kob-Andersen mixture when applying \textit{NVU} dynamics on the hypersphere model. \textit{NVU} dynamics is defined\cite{IngebrigtsenTrondS2011NdIG} by geodesic motion on $\Omega$ and has been shown\cite{IngebrigtsenTrondS2011NdIC} to be equivalent to \textit{NVE} dynamics in the thermodynamic limit.

A second motivation for studying the hypersphere model originates in the theory of strongly correlating liquids. It has been shown\cite{GnanNicoletta2009Pcil} that the reduced-unit constant-potential-energy hypersurface (denoted $\tilde \Omega$) of a strongly correlating system is invariant along the so-called isomorphs in the density-temperature thermodynamic phase diagram, which are defined as curves of constant excess entropy (the configurational adiabats). Along these curves, static and dynamic correlation functions are invariant when expressed in reduced units\cite{GnanNicoletta2009Pcil}. In a strongly correlating system (an ``R-simple'' system), the virial and the potential energy correlate better than 90\% in their thermal-equilibrium fluctuations in the \textit{NVT} ensemble\cite{IngebrigtsenTrondS2012WIaS}. Generally, van der Waal bonded and metallic liquids are strongly correlating \cite{gun11,HummelFelix2015Hsio}. 

As shown in Ref. \onlinecite{GnanNicoletta2009Pcil} all strongly correlating liquids have isomorphs to a good approximation. There it is also shown that the existence of isomorphs in the phase diagram, along which the $\tilde \Omega$ hypersurface is (almost) invariant, implies that for any strongly correlating liquid there exists a single-parameter family of $\tilde \Omega$ hypersurfaces. It has moreover been argued\cite{DyreJeppeC2013Npos} that the family of $\tilde \Omega$ hypersurfaces is approximately the same for all strongly correlating simple liquids. Thus, there might exist a one-parameter model of a constant-potential-energy hypersurface, from which the dynamics of most or all R-simple liquids can be derived. Our conjecture is that the hypersphere model might approximate such a surface.

We use a slightly modified version of the hypersphere model proposed by Samuelsen \textit{et al.} by letting the dynamics be defined by a random walk on the \underline{inside} of the hypersphere model. Thus we redefine $\Omega$ to be the part of configuration space considered, 

\begin{equation}\label{eq:PELE2}
        \Omega=\{ (\textbf{r}_1,...,\textbf{r}_N)\in \mathbb{R}^{3N}\ |\ U(\textbf{r}_1,...,\textbf{r}_N) \leq U_0\}.
\end{equation}
This is similar to that of the potential energy landscape ensemble in Refs. \onlinecite{WangChengju2007Gpot,WangChengju2007Gpot2} (with a different dynamics). Increasing the number of dimensions of the hypersphere system forces the random walk towards the surface of the hyperspheres, thus approximating a random walk on the surface of the hypersphere model. Using an upper limit makes things much simpler computationally, however, since no additional calculations are needed to guarantee the random walker actually is on the surface.

Through theoretical considerations of an ordinary potential energy landscape we construct in Sec. \ref{sec:theModel} the hypersphere model. As the number of dimensions increases, the ratio between the volume of a hypersphere of radius $r$ and a hypercube with sidelength $2r$ decreases at such a pace that too many spheres are needed in order to fill a hypercube to the required densities, thus rendering it impossible to simulate in high dimensions. In order to circumvent this problem - to a certain degree - we will in Sec. \ref{sec:sphereGen} present an algorithm for generating spheres on the fly. The algorithm is similar to that of Ref. \onlinecite{BreretonTim2014Acef} by generating spheres only when and where they are needed for the random walks. This also means that we avoid the use of periodic boundary conditions and thus any finite-size effects caused by these. Using this algorithm, we performed random walks in the 12-, 18-, 30-dimensional hypersphere model. Details about the random walks and its inherent dynamics are described in Sec. \ref{sec:RW}. Since the aim is to study diffusion at time scales way beyond what is possible using random walks, we in Sec. \ref{sec:NfoldKinetic} implement inherent dynamics using the kinetic Monte Carlo method\cite{BulnesF.M1998Csdn}. We also implement two other optimization methods and show that the resulting inherent dynamics simulations provide similar results to that of the random walks in the 18- and 30-dimensional hypersphere model. In Sec. \ref{sec:Results} we present the results of the kinetic inherent dynamics simulations. Here, we study the diffusion coefficients in the 18-, 21-, 24-, 27- and 30- dimensional hypersphere model. It is also shown that time-temperature superposition is satisfied at low densities (corresponding to low temperatures) in the 33-, 42- and 45-dimensional hypersphere model, and that these are also well fitted by the RBM\cite{SchroderTB2008PRL} and thus also to the experimental data highlighted in Ref. \onlinecite{BierwirthSPeter2017GPMR}. Finally, we discuss some problems of the model and why, even given those problems, the hypersphere model and the RBM provide similar results at low densities.

\section{\label{sec:theModel}The hypersphere model}

Local minima of the potential-energy function are called inherent structures.\cite{Stillinger1935}  The set of configurations in the potential energy landscape that maps to a given inherent structure under a steepest gradient descent is called the basin of attraction of the inherent structure in question.\cite{Stillinger1935} It was stipulated by Goldstein\cite{GoldsteinMartin1969VLat} that below a certain crossover temperature, liquid dynamics is mainly governed by vibrations interspersed by transitions between basins of the potential-energy landscape. Numerical evidence of this was later found by Schr{\o}der \textit{et al.}\cite{SchroderTB2000Ctpe} by the use of Newtonian Molecular Dynamics. 

At low temperatures, it has been observed\cite{SastrySrikanth1998Sodd} that the mean squared distance between a typical liquid configuration and its corresponding inherent structure to a good approximation is consistent with harmonic vibrations around the inherent structure.
The potential energy at a typical low temperature liquid configuration can therefore be approximated by a harmonic potential:
\begin{equation}\label{eq1}
    U(x_1,x_2,...,x_{3N}) \approx U_\mathbf{q} + \sum_{i,j=1}^{3N}\frac{1}{2}c_{\textbf{q},ij}(x_i-q_i)(x_j-q_j)
\end{equation}
in which $\mathbf{q} = (q_1,q_2,..., q_{3N})$ are the coordinates of the inherent structure and $U_\mathbf{q}$ its potential energy. Inserting the harmonic potential $U$ into Eq. (\ref{eq:PELE2}) yields a hyperellipsoid centered at $\mathbf{q}$ in configuration space. 

In order to arrive at the hypersphere model, we impose three properties on the system, which all are significant simplifications compared to the potential-energy landscape of, for example, the Kob-Andersen binary Lennard-Jones liquid:
\begin{enumerate}
    \item The potential energy around the inherent structures is assumed to be isotropic ($c_{\textbf{q},ij}$ in \eq{eq1} proportional to the identity matrix, for each $\textbf{q}$). Thus, the hyperellipsoids become hyperspheres.
    \item The curvature (determined by the $c_{\textbf{q},ij}$'s) and the potential energy at the inherent structures are assumed to be the same for all the inherent structures across the potential energy landscape. That is $U_\textbf{q}$ and $c_{\textbf{q},ij}$ each have identical values at all inherent structures. This means that all spheres have the same radius.
    \item Inherent structures are assumed to be randomly distributed in configuration space, without any correlations.
\end{enumerate}
By imposing these properties, the model becomes a one-parameter model since the only parameter is the reduced density of hyperspheres. The reduced density is defined according to Ref. \onlinecite{TorquatoS2012Eodo}: If $V_d(r)$ is the volume of a sphere of radius $r$ in $d$ dimensions, the reduced density $\eta$ is given by
\begin{equation}
    \label{eq:redDens}
    \eta = \varrho V_d(r)
\end{equation}
\noindent with $\varrho$ being the number density of spheres in $\mathbb{R}^d$. Physically, the reduced density is the sum of all sphere volumes divided by the total configuration-space volume. Note that due to overlaps this is not the volume fraction, $\phi$, of the hypersphere system. These two numbers are related by
\begin{equation}
    \phi = 1-e^{-\eta}
    \label{eq:vol_frac_red_dens}
\end{equation} 
in any number of dimensions\cite{alma99122654779705763}. Notice that $\eta \ll 1$ implies $\phi\approx\eta$. Since we mainly simulate at very low densities ($\eta<0.01221$), the difference between $\phi$ and $\eta$ becomes negligible. We avoid simulating too low densities, i.e., near the percolation threshold, since diffusion requires a percolating cluster of hyperspheres.

\section{\label{sec:sphereGen}Sphere-generating algorithm}

\begin{figure*}
\includegraphics[width=\linewidth]{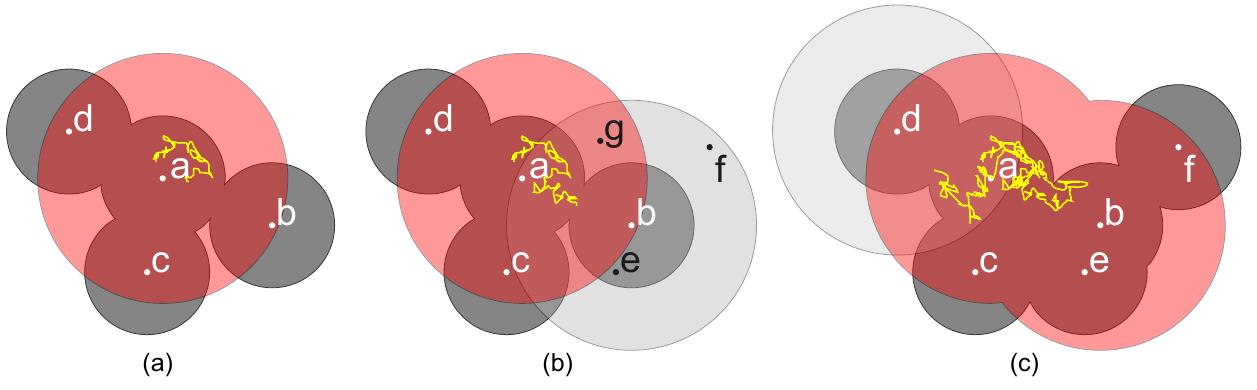}
\caption{Sphere-generating algorithm. Spheres are represented by dark grey circles, and random walks are represented by the yellow line. 
{(a)} Since the random walk starts from a random point in sphere a, all its neighbors have already been determined (spheres $b$, $c$ and $d$). The centers of these were all randomly distributed in a sphere of radius $2r$ (the red area). 
{(b)} The random walk has just entered sphere $b$ for the first time. Random neighboring spheres are now determined for sphere $b$, leading to points $e$, $f$, and $g$. 
{(c)} Since sphere $a$ already has had all its neighbors determined, the point $g$ is removed. The spheres $e$ and $f$ are then added to the system of spheres, and the random walk is allowed to continue until it reaches a previously unvisited sphere (sphere $d$), at which point the procedure is repeated.}
\label{fig:rwIl}
\end{figure*}

The simplest way to generate a system of hyperspheres is to randomly distribute spheres inside a hypercube with periodic boundary conditions. If the hypercube is too small, however, the random walk might be able to enter the same neighbouring sphere from two or more different places of the current sphere. To avoid this problem, a cube length greater than $4r$ is needed ($r$ is the radius of the spheres). Thus, just to reach the percolation threshold $\eta_c$ (which approaches $2^{-d}$ from above as the number of dimensions increases\cite{TorquatoS2012Eodo}), a minimum of $H$ spheres are required:
\begin{equation}
     \eta_c \approx \frac{1}{2^d} = \frac{H V_d(r)}{(4r)^d} \Leftrightarrow     H = 4^d \frac{\Gamma(\frac{d}{2}+1)}{\pi^\frac{d}{2}}\frac{1}{2^d}\,.
\end{equation}
Here, $\Gamma$ denotes the Gamma function. In 30 dimensions this amounts to more than $10^{13}$ spheres, which is computationally impossible to achieve. In order to circumvent this, we use an algorithm similar to that of Ref. \onlinecite{BreretonTim2014Acef} that generates spheres on-the-fly. Appendix \ref{appendix:Proof} demonstrates that the algorithm works as intended, i.e., it generates a system of spheres centered on randomly and independently distributed points of point density $\varrho$ within the desired volume (a Poisson point process).

The algorithm works by only adding spheres to the subset of $\mathbb{R}^d$ that is relevant for the random walk (the red and light grey areas in Fig. \ref{fig:rwIl}). The only spheres that are relevant are the ones that have been visited during the random walk and all of their neighbours. Since the centers of overlapping spheres can be no more than $2r$ apart, the relevant subset of $\mathbb{R}^d$ must be a union of $d$-balls of radius $2r$. Each time the random walk enters a new sphere that has not previously been visited, the set of neighbours to this sphere is determined once and for all - all being in the $d$-ball of radius $2r$ centered at the newly visited sphere. This means that spheres can neither be added nor removed to/from this volume after the first visit (\fig{fig:rwIl}). Since all spheres have been assigned to this volume, we call this volume the assigned volume.

In order to add neighbors to a new sphere, we utilize the fact that the number of neighbours to any given sphere is Poisson distributed with mean $2^d\eta$ (see Appendix \ref{sec:ConsSigma2}). Pick a number $k$ from this distribution and place $k$ random points in a $d$-ball of radius $2r$ (\fig{fig:rwIl}(b)). By removing the points, e.g., point $g$ in \fig{fig:rwIl}(b) that lies in the assigned volume, we can ensure that the number of neighbors to each sphere is still Poisson distributed with mean $2^d\eta$. That this is the case is shown in Appendix \ref{appendix:Proof}. A more detailed description of the algorithm is also given in the Appendix. Using this algorithm the relevant subset of $\mathbb{R}^d$ expands each time the random walk enters a sphere that has not been visited before.

\section{Random walks}\label{sec:RW}

\begin{figure}
    \centering
    \includegraphics[width=.95\linewidth]{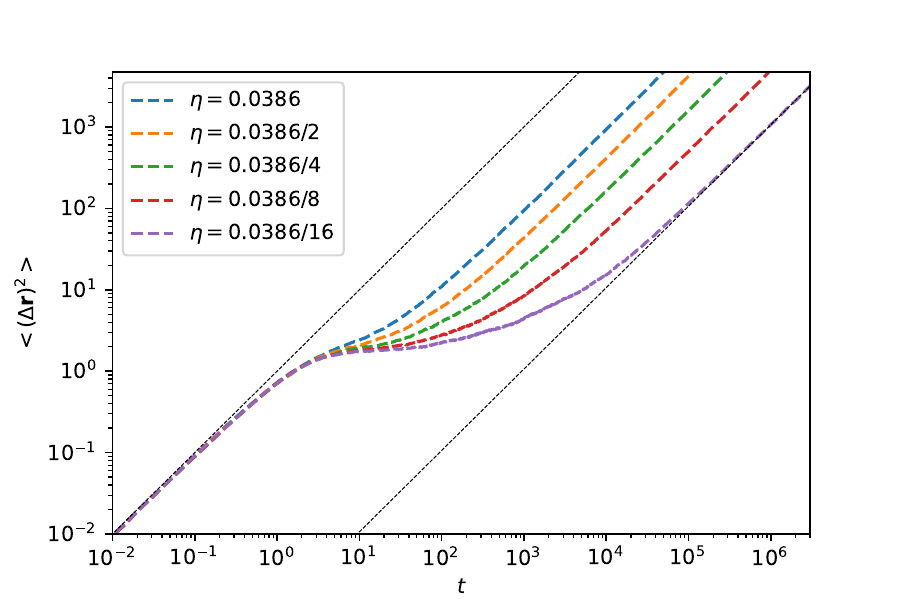}
    \caption{The MSD for random walks in the 12-dimensional hypersphere model at five different reduced densities. Each curve is averaged over 1000 random walks of step length $a= 0.02$. It can be seen that the dynamics get slower as the reduced density is decreased, which is a general behavior for the model. The black dashed lines have slope unity, corresponding to diffusive behavior.}
    \label{fig:RW_12D}
\end{figure}

The random walks were initiated from a random point inside a hypersphere (that already had its neighbors generated). If a walker tried to exit the boundary of the current sphere, it was checked whether the random walker entered a neighbouring sphere. If so, neighbours were generated to the new sphere as described above before allowing the random walk to continue (\fig{fig:rwIl}). If not, the step was not performed but the time counter was still increased. In order to equilibrate the systems, random walks were performed for the same amount of time as the production runs. 

The results of the random walk should be independent of step length and only depend on time. In order to achieve this, we note that the MSD for a random walk in $\mathbb{R}^d$ is given by
\begin{equation}
    \label{eq:RandWalkInR3}
    <(\Delta \mathbf{r})^2>=Na^2
\end{equation}
where $N$ is the number of steps and $a$ is the step length. 

By defining the time of the random walk as being proportional to the number L defined by $N=L/a^2$, the random walk as a function of time becomes independent of the step length. For a random walk in a system of hyperspheres, at short time scales one expects behavior similar to that of a random walk in $\mathbb{R}^d$. It therefore seems reasonable that the time after $i$ random walk steps should be given
\begin{equation}
    t(i) = ka^2i
    \label{eq:TimeScale}
\end{equation}
where $k$ is the constant of proportionality between the time and $L$. For all random walks we put $k=1$.

In \fig{fig:RW_12D} we show the MSD for a random walk in the 12-dimensional hypersphere model at five reduced densities. The dynamics slow down in a manner similar to what is typically observed for supercooled liquids: at long times there is a diffusive regime, separated from the short-time dynamics by a plateau that grows in extent with decreasing reduced hypersphere density, corresponding to decreasing temperature. The relationship between reduced density and temperature is discussed further below. 

\begin{figure}
    \centering
    \includegraphics[width=.9\linewidth]{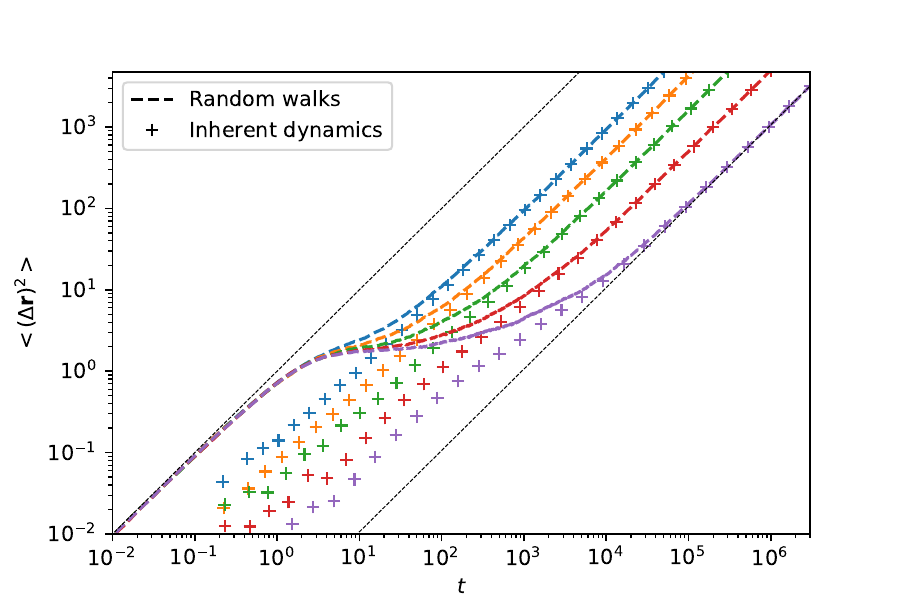}
    \caption{Comparison between the MSD for the random walk and its corresponding inherent dynamics in the 12-dimensional hypersphere model. The black dashed lines have slope unity.}
    \label{fig:RW_12D_ID}
\end{figure}

\fig{fig:RW_12D_ID} shows results from the same simulations as in \fig{fig:RW_12D}, adding the so-called inherent dynamics \cite{SchroderTB2000Ctpe}. This is a mapping, where configurations are mapped to the corresponding inherent structure before computing dynamical properties, here the MSD. Note that this is only done when calculating the MSD - not when performing the random walks. The effect of this procedure is to remove the effect of the vibrations around inherent states, and the MSD of the inherent dynamics will therefore only relay information about transitions between basins. Since the center points of the spheres correspond to the inherent structures, the inherent dynamics of the random walk is easily found in the hypersphere system. 




\section{\label{sec:NfoldKinetic}Kinetic Monte Carlo inherent dynamics}

\begin{figure}[b]
\includegraphics[width=.7\linewidth]{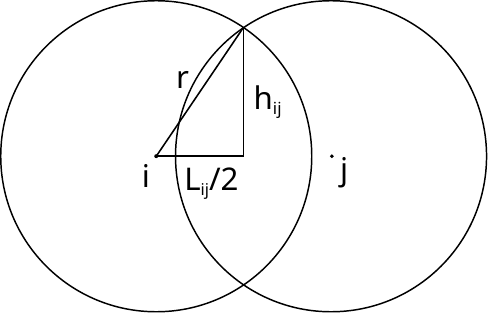}
\caption{\label{fig:probability_of}Illustration of the radius $h_{ij}$ of the intersecting $(d-1)$-dimensional sphere between the two neighbouring spheres $i$ and $j$.}
\end{figure}

A given sphere has a finite number of neighbors. This gives us the possibility of simulating the inherent dynamics directly, using the kinetic Monte Carlo simulation method\cite{BulnesF.M1998Csdn}. Instead of waiting for a transition to happen during a random walk, the kinetic Monte Carlo method draws a time to transition $\Delta t$ using the probability $W_i$ of jumping from sphere $i$ to one of its neighbours. To determine $W_i$ we let $L_{ij}$ denote the distance between two neighbouring $d$-dimensional spheres, $i$ and $j$. The radius $h_{ij}$ of the "intersecting" $(d-1)$-dimensional sphere is given by (\fig{fig:probability_of}). 
\begin{equation}
    h_{ij}=(r^2-L_{ij}^2/4)^{1/2}\,.
    \label{eq:h}
\end{equation}

As shown in Appendix \ref{appendix:Rates}, for the relevant values of $h_{ij}$, the probability of transitioning from sphere i to sphere j is given by 
\begin{equation}
\Gamma_{ij} = \Gamma_0 L_{ij} \left(h_{ij}/r\right)^{d-2}\,.
\label{eq:jump_probability}
\end{equation}
The constant of proportionality $\Gamma_0$, which only depends on the number of dimensions, can be found by comparing the MSD for the inherent dynamics of the random walk and the kinetic method at a single reduced density; $W_i$ is then defined by
\begin{equation}
W_i = \sum_{k} \Gamma_{ik},
\label{eq:suuum}
\end{equation}
where $k$ are the neighbours of sphere $i$. It should be noted that this probability does not take overlap between intersecting $(d-1)$-dimensional spheres of neighbours into account. However, due to the low densities (see the relation between reduced density $\eta$ and volume fraction $\phi$ in Eq. \ref{eq:vol_frac_red_dens}) and high number of dimensions, we assume that these overlaps are few and have minimal effect on the results. 

The time to transition $\Delta t$ is given by
\begin{equation}
    \Delta t = -\frac{1}{W_i}\ln{\zeta}
\end{equation}
where $\zeta$ is randomly uniformly distributed between 0 and 1.\cite{BulnesF.M1998Csdn} 

The following further optimizations were implemented. First, the previously visited spheres and their neighbors are deleted along the way. Unless otherwise mentioned, only the last 1000 unique previously visited spheres are stored. Their neighbors are also stored in a corresponding neighbor list. The neighbor list of the last 999 unique previously visited spheres can then be used to create a complete list of spheres when a new sphere is entered. Combining this with the spheres generated by the sphere-generating algorithm completes the neighbours for the newly visited sphere and ensures that the neighbours for all of the 1000 stored unique visited spheres remain constant. Secondly, due to the increasing amount of jumping back and forth between the spheres when lowering the density toward the percolation threshold, we recall (instead of recalculating) the jump probabilities, $W_i$, etc., when reentering a previously visited sphere. These optimizations are only important at low densities. In fact, the recalling optimization results in longer simulation time at the higher densities, since the same spheres are rarely visited many times. 
\begin{figure*}
    \centering
    \includegraphics[width =.95\linewidth]{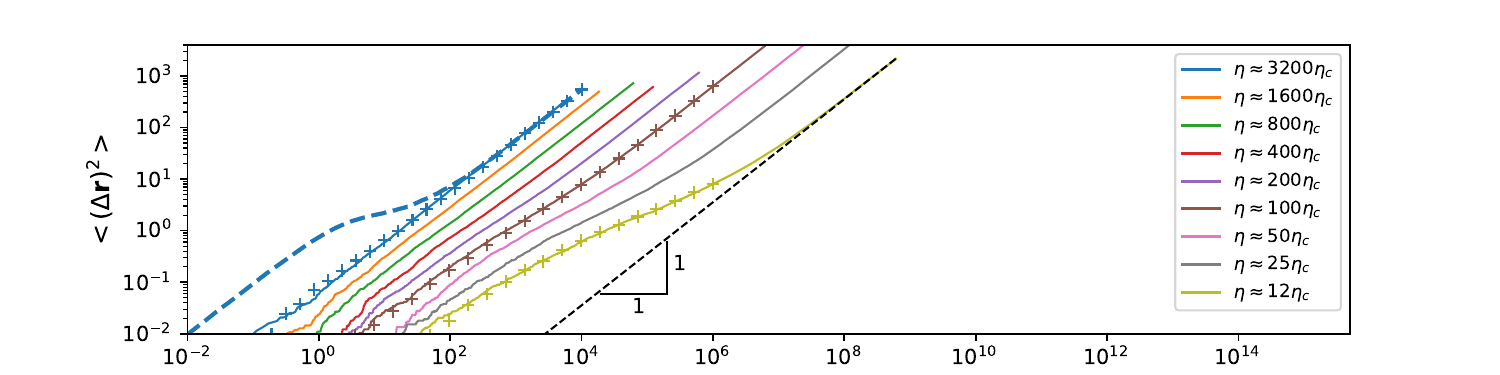}
    \caption{Comparing the MSD of the inherent dynamics using the random walk (pluses) and the kinetic Monte Carlo method (full lines) in the 18-dimensional hypersphere model. The dashed blue line represent the random walk itself at the highest simulated density.}
    \label{fig:D18}
\end{figure*}

In \fig{fig:D18} we show that the MSD of the inherent dynamics of the random walk and the fully optimized kinetic Monte Carlo inherent dynamics provide similar results in the 18-dimensional hypersphere model\footnote{Except for the Kinetic Monte Carlo inherent dynamics being a bit faster than the random walk at the highest density, it is evident that these dynamics are equivalent at longer time scales. We hypothesize that the difference at the highest density arises because of significant overlap of neighbors in the current hypersphere, thus causing $W_i$ to be larger than it should be and causing faster dynamics. We have therefore scaled the time so that the curves overlap at the lowest density simulated.}. The reduced densities are determined in terms of the average number of neighbours $\langle N_b \rangle$  to a single sphere (see Eq. \ref{eq:meanNeighbours}):
\begin{equation}
    \langle N_b \rangle = 2^d\eta \Leftrightarrow \eta = \langle N_b \rangle2^{-d} \approx \langle N_b \rangle\eta_c,
    \label{eq:neighbour_percThres_relation}
\end{equation}
Recall that $\eta_c$ also denotes the percolation threshold.

\begin{figure*}
    \centering
    \includegraphics[width = .95 \linewidth]{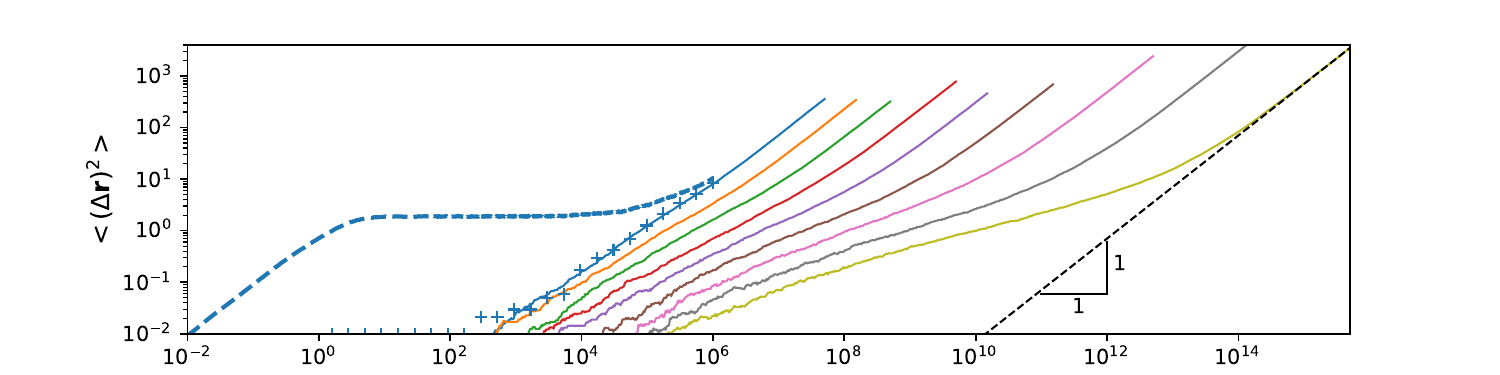}
    \caption{Comparing the MSD of the inherent dynamics using the random walk and the kinetic Monte Carlo method in the 30-dimensional hypersphere model. The reduced densities have the same proportionality to the 30-dimensional percolation threshold as the 18-dimensional percolation thresholds in \fig{fig:D18}, i.e., the average number of neighbours are the same for the same coloured lines.}
    \label{fig:D30}
\end{figure*}

In \fig{fig:D30} we show the MSD as a function of time for the inherent dynamics of the 30-dimensional hypersphere model at very low reduced densities, corresponding to low temperatures in a real system\footnote{In order to compare the random walk with the optimized kinetic Monte Carlo version at $\eta\approx 3200\eta_c$ for $n=30$, we had to reduce the number of stored uniquely visited spheres (and their neighbours) to $50$, due to the increasing simulation time at higher densities. For the next four densities, we used a version of the code with no additional optimizations other than the kinetic Monte Carlo method. At even lower densities, the fully optimized code became faster.}. It is possible to run simulations in higher dimensions, but here it becomes harder to attribute a time scale to the simulations. This is due to the lack of knowledge of how $\Gamma_0$ scales with the number of dimensions. The average number of neighbours to a single sphere creates a practical upper bound on the reduced density depending on the number of dimensions. Likewise, the lower the reduced density, the longer a random walk is needed in order to compare the results with the kinetic Monte Carlo method. Thus, there exists a natural limitation to the number of dimensions at which we can compare diffusion coefficients for the random walks. 

\section{Results}\label{sec:Results}

\subsection{Diffusion coefficients}
\begin{figure}
    \centering
    \includegraphics[width=.9\linewidth]{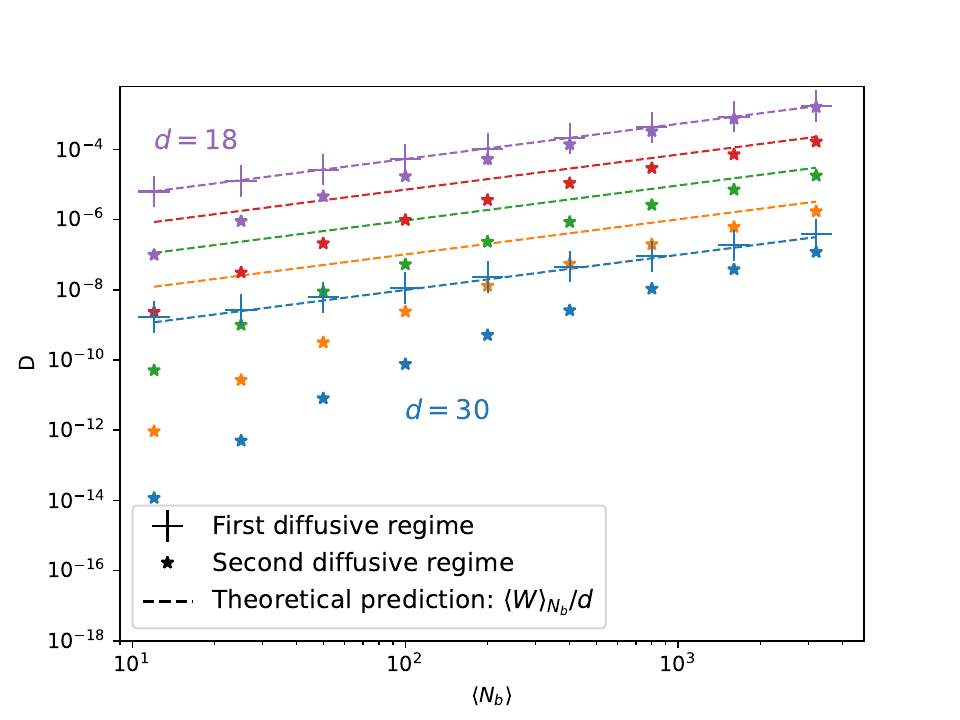}
    \caption{Diffusion coefficients and the average transition rate from any sphere in the system with $N_b$ average neighbours vs. average number of neighbouring spheres in the 18- (purple), 21- (red), 24- (green), 27- (orange) and 30-dimensional (blue) hypersphere model.}
    \label{fig:initDiffusion}
\end{figure}

In \fig{fig:initDiffusion} we plot the diffusion coefficient $D$ as a function of the average number of neighbors $\langle N_b \rangle$ for $d=18,21,24,27,30$.

A lower bound of the diffusion coefficient can be established by finding the average transition rate from a random sphere in the system.  We do this by noticing that the distance between the centers of any two neighboring spheres is given by $L_{ij}=2rS^{1/d}$ (see Eq. (\ref{eq:randradius})), where $S$ is a random variable that is uniformly distributed in [0,1]. Combining this with Eqs. (\ref{eq:h}) and (\ref{eq:jump_probability}) yields
\begin{equation}
    \frac{\Gamma(S)}{\Gamma_0} = 2rS^{1/d}(1-S^{2/d})^{\frac{d-2}{2}}\,.
\end{equation}
Given that a sphere has $N_b$ neighbours, the probability of transitioning from the sphere into any of its neighbours is given by:
\begin{equation}
    W = \sum_{i=1}^{N_b}\frac{\Gamma(S_i)}{\Gamma_0}\,.
\end{equation}
Thus, the average probability of transitioning from a sphere with $N_b$ neighbours into any of its neighbours is given by
\begin{align}
    \langle W \rangle_{N_b} &= \int_0^1...\int_0^1\sum_{i=1}^{N_b}\frac{\Gamma(S_i)}{\Gamma_0}dS_1,...dS_{N_b}\nonumber \\
    &= \frac{N_b}{\Gamma_0}\int_0^12rS^{1/d}(1-S^{2/d})^{\frac{d-2}{2}}dS
    \label{eq:FirstDiffusiveRegime}
\end{align}
Since the number of neighbours are Poisson distributed and the integral is independent of the number of neighbours, this is also the average transition rate from any sphere in the system. Plotting $\langle W \rangle_{N_b}/d$ alongside the diffusion coefficients yields  
\fig{fig:initDiffusion}. In order to achieve the results for the first diffusive regime of the inherent dynamics, we averaged over one million short-time simulations; $\langle W \rangle_{N_b}/d$ is indeed the diffusion coefficient for the first diffusive regime.

We now proceed to discuss how to relate the reduced density, $\eta$, to the temperature, $T$, considering the potential energy surface of an atomic glass-forming liquid as the starting point. For a given state point, the hypersphere model is arrived at by assuming the potential energy in the basin of attraction of the relevant inherent states is given by d-dimensional harmonic oscillators with identical spring constants $k$. From the equipartition theorem it then follows that $r^2 = d k_BT/k$. Combining this with Eq.~(\ref{eq:redDens}), and using that $V_d(r) = V_d(1)r^d$, yields
\begin{equation}
    k_BT = \frac{k}{d}\left(\frac{\eta}{\varrho V_d(1)}\right)^{2/d}\,.
    \label{eq:temperature}
\end{equation}
From this, we define a reduced temperature, $\tilde T$, which is proportional to the real temperature for a fixed energy landscape in the sense that $\varrho$ and $k$ are independent of state point:
\begin{equation}
    \tilde T = \frac{1}{d}\left(\frac{\eta}{V_d(1)}\right)^{2/d} = \frac{\varrho^{2/d}}{k}k_BT\,.
\end{equation}

\begin{figure}
    \centering
    \includegraphics[width=.9\linewidth]{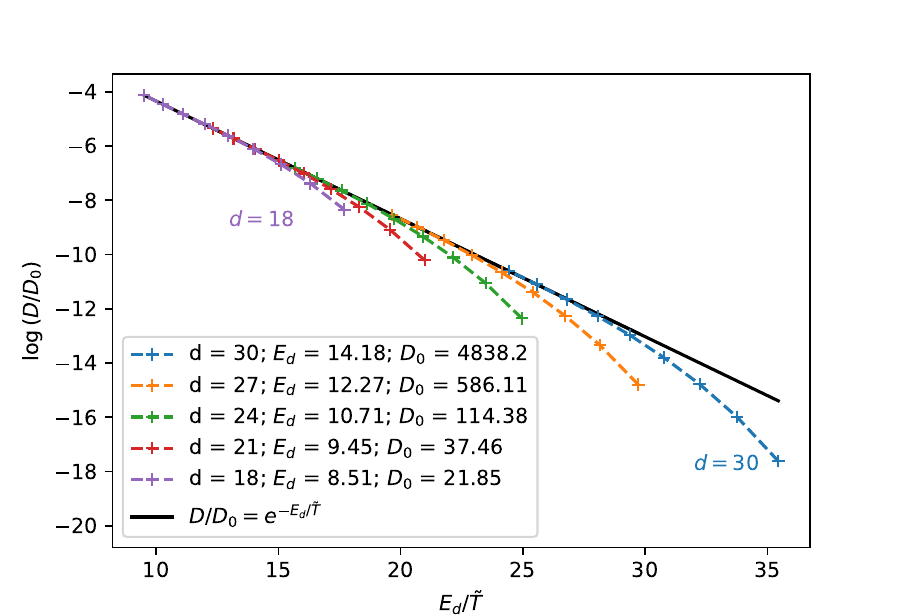}
    \caption{``Arrhenius'' plot of the scaled diffusion coefficients.}
    \label{fig:ArrheniusPlot}
\end{figure}

In Fig. \ref{fig:ArrheniusPlot} the diffusion coefficients are plotted in an ``Arrhenius'' plot in terms of the reduced temperature. The applied scaling parameters are found from fitting the high-temperature results to Arrhenius behavior $D(\tilde T)=D_0e^{-E_d/\tilde T}$. We observe an apparent approach to Arrhenius behavior as the number of dimensions increases. Expressing this in terms of the non-reduced temperature (see Eq. (\ref{eq:temperature})), leads to,
\begin{equation}
    D(T)=D_0e^{-\frac{kE_d}{\varrho^{2/d}k_BT}}\,.
\end{equation}

We stress that the observed (apparent) Arrhenius behavior is found when varying temperature in a fixed energy landscape, i.e. for constant $k$ and $\varrho$. 
We leave a detailed analysis of how to map the model back to real systems to future investigations, but note that both $k$ and $\varrho$ in general depend on state point. In particular, when a liquid is cooled, lower parts of the potential energy landscape is explored\cite{SastrySrikanth1998Sodd}, and the number of relevant inherent states decrease \cite{BuchnerStephan1999Pelo}, i.e., the number density $\varrho$ decreases. Via Eq. (\ref{eq:temperature}) this results in an apparent activation energy that increases with decreasing temperature, i.e., the non-Arrhenius behavior that is a hallmark of viscous liquids.

\begin{figure*}
    \centering
    \includegraphics[width = \linewidth]{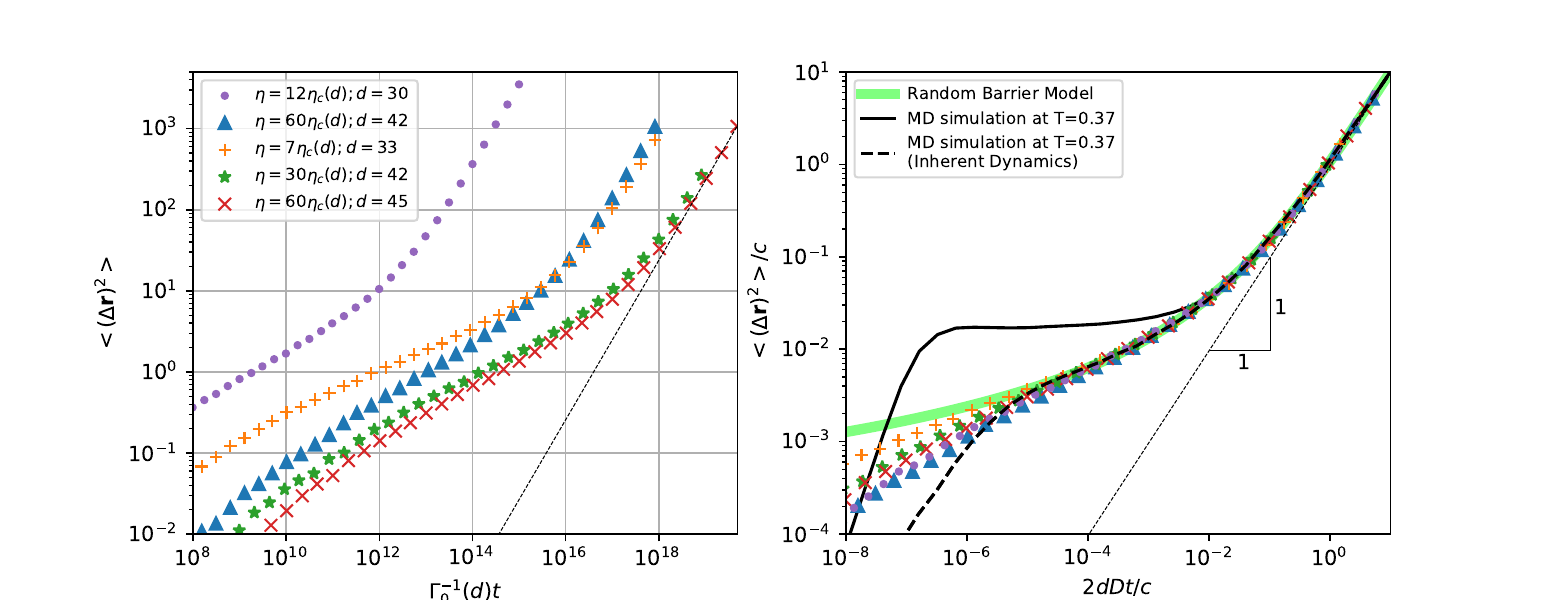}
    \caption{(a) The MSD of the kinetic Monte Carlo inherent dynamics at reduced densities nearing the percolation threshold. 
    (b) Data from (a) scaled to fit the RBM prediction \cite{SchroderTB2008PRL}. In addition to these data, we have also plotted the MSD from simulations\cite{SchrderThomasB2020Smdi} of a crystallization-resistant version of the Kob-Andersen binary Lennard Jones mixture at $T=0.37$.}
    \label{fig:rbm}
\end{figure*}

\subsection{The Random Barrier Model}
When the actual time is not needed such as when scaling onto a master curve, we can ignore $\Gamma_0$ and thus simulate at a higher number of dimensions than 30. We have therefore run simulations of the 33, 42 and 45-dimensional hypersphere model. In \fig{fig:rbm} we fitted the data to the MSD of the RBM\cite{SchrderThomasB2020Smdi}. The figure shows that these give similar results to those achieved by molecular dynamics simulations\cite{SchrderThomasB2020Smdi}. It can be seen that increasing the number of dimensions and/or decreasing the density brings the results closer to the prediction of the RBM. Due to the similarities between the models, this should not come as a surprise. The RBM is a cubic lattice model with identical site energies and random energy barriers for nearest-neighbor jumps\cite{DYREJ.C1988Trfb, SchroderTB2008PRL}. The latter two properties are also present in the hypersphere model. The conjecture that lattice and continuum percolation are in the same universality class\cite{GawlinskiET1981Cpit,MertensStephan2012Cpti,alma99122654779705763} could thus explain why the MSD as a function of time for the inherent dynamics in the hypersphere model approaches that of the RBM prediction at low densities. 

\section{Discussion}

In this paper a method is presented that makes it possible to generate simple models of NVU surfaces on-the-fly (or surfaces capped by a potential energy) with randomly distributed inherent structures. In order to do this, only knowledge of the shape of the basins in configuration space is needed. Knowing the transition rates between the basins (using any type of dynamics) allows fast simulations of inherent dynamics, since vibrational motion is ignored and no energy minimization is required. 

For the 18 dimensional hypersphere model, we performed inherent dynamics simulations of random walks up to $5\cdot10^{15}$ MD time units. This corresponds to hundreds of seconds in argon units. We also showed that the average behavior of a random walk several decades prior to diffusion is described by the RBM. This encapsulates the behavior of real glass-forming liquids, compare the experimental results of Bierwirth \textit{et. al}\cite{BierwirthSPeter2017GPMR}. 

Although the agreement with the typical viscous liquid behavior is striking in view of the simplicity of the model, there are outstanding issues that deserve further investigation. In particular, we point out the following: Consider the results for the 30-dimensional hypersphere model with $\eta=12\eta_c$ in \fig{fig:rbm}. These results are in good agreement with the RBM for about the same range of reduced times as the Kob-Andersen model. However, the duration of the plateau in MSD for the Kob-Andersen model is approximately 4 decades. Referring to \fig{fig:D30} we see that for the 30-dimensional hypersphere model, a similar duration of the MSD plateau is found for $\eta=3200\eta_c$ (blue curve) where there is hardly any hint of RBM-behavior; the inherent dynamics is close to simply being an extrapolation of the diffusive behavior. We conclude from this that much of the slowing down is due to the system being localized in a single hypersphere, which gives a plateau in the MSD but no contribution to the inherent MSD. Whether this is due to all inherent structures having the same potential energy, the shape of their basins (hyperspheres, instead of the more general hyperellipsoids), their completely random and independent distribution, or simply the random walk dynamics itself, is an interesting open question for future research. 

\begin{acknowledgments}
This work was supported by VILLUM Foundation grants VIL16515 and VIL23189.
\end{acknowledgments}
\newpage
\appendix

\section{\label{appendix:Proof}Validating the sphere generating algorithm.}

\begin{figure}
    \centering
    \includegraphics[width=\linewidth]{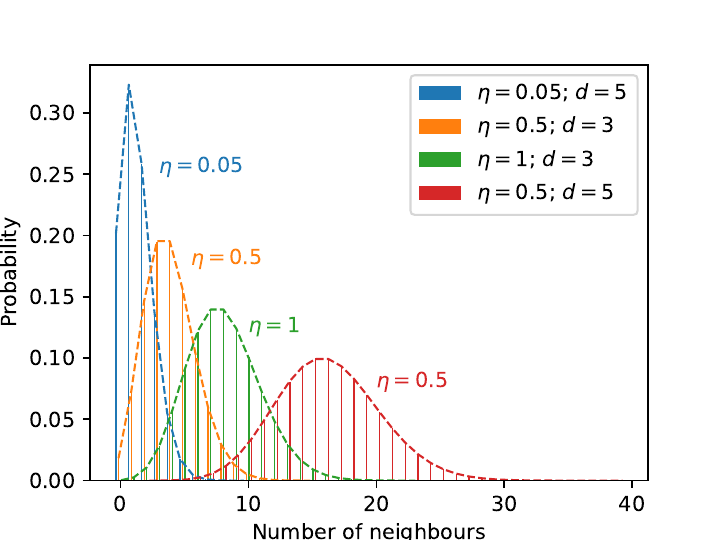}
    \caption{For each ($\eta,d$) 1000 $d$-dimensional cubes with periodic boundary conditions were made. The cube length was decided such that there would be on average 200 spheres of radius 1 inside the cube. A $d$-dimensional integer lattice of points were generated and spheres were placed around these using the assigned volume principle of the algorithm. The dashed lines marks the Poisson distribution with mean $2^{d}\eta$, whereas the bars lines show the neighbour count.}
    \label{fig:TestingNeighGen1}
\end{figure}

Let $\mu$ denote a countable set of independently and randomly distributed points in $\mathbb{R}^d$ of point density $\varrho$, which is by definition a homogeneous Poisson point process on $\mathbb{R}^d$ with density $\varrho$.
 
Let $\bar B^d_{\mathbf{q},r}$ denote a closed $d$-ball of radius $r$ centered at $\mathbf{q}$, i.e., 
\begin{equation}
   \bar B^d_{\textbf{q},r} = \{\textbf{x}\in\mathbb{R}^d||\textbf{q}-\textbf{x}|\leq r\}.
\end{equation}
Then the $d$-dimensional hypersphere model defined by $\mu$, is defined as the surface of the union 
\begin{equation}
    \bigcup_{\mathbf{q}\in \mu}\bar B_{\mathbf{q},r}^d.
\end{equation}

A model consisting of a system of hyperspheres, whose centers constitute a homogeneous Poisson point process on $\mathbb{R}^d$, is referred to in the mathematical literature as the Poisson "blob" model (among other names); it is also one of the simplest examples of a Boolean model, cf. \cite{PenroseMathewD1991Oacp}, \cite{Stoyan1}. For a given realization of the model, the connected components of the union of hyperspheres are called \emph{clusters}. The number of spheres in a given cluster, is called the size of the cluster.

We start this section by presenting two numerical tests, that show that the method presented in section \ref{sec:sphereGen} works as intended. First we show that the number of neighbours to an arbitrary point in $\mu$ (equivalently an arbitrary sphere in the model) is Poisson distributed with the correct mean, followed by showing that the algorithm grows clusters of the correct size distribution according to theory, see also sections \ref{secconditional} and \ref{secproof}. 

The algorithm generates a realization of a Poisson point process of point density $\varrho$ on-the-fly, i.e. the algorithm generates the points, and hence the spheres in the hypersphere model, step by step as they are needed. 

A more precise definition of a Poisson point process will follow after a more detailed description of the algorithm. We will thereafter show by induction that, indeed, a realization of a Poisson point process is formed when using this algorithm. 

\subsection{Testing the algorithm numerically}

In order to show that the method of section \ref{sec:sphereGen} provides a Poisson distributed number of neighbours with mean $2^d\eta$, we performed tests where we ensure that the entire volume of a $d$-dimensional cube is assigned by placing spheres using lattice points (not regarded as spheres themselves). These lattice points are along each axis separated by distance 1. Spheres are then added for each lattice point, according to the method described in section \ref{sec:sphereGen}. 

In order to ensure, that the entire volume is assigned, we need to guarantee that the spheres have a large enough radius. Since we place the actual spheres using "fictional spheres" of radius $2r$, we need to ensure that these "fictional spheres" fill out the entire volume. We do this by noticing that the distance from any lattice point to the center of a hypercube of side length 1 is given by $(0.5^2d)^{1/2}$. Thus, we require that $2r>(0.5^2d)^{1/2}$. For the results in Fig. \ref{fig:TestingNeighGen1}, we used $r=1$. 

For each of the spheres placed, the number of neighbours were counted. It can indeed be seen that the number of neighbors is Poisson distributed with mean $2^d\eta$.



\begin{figure}
    \centering
    \includegraphics[width=\linewidth]{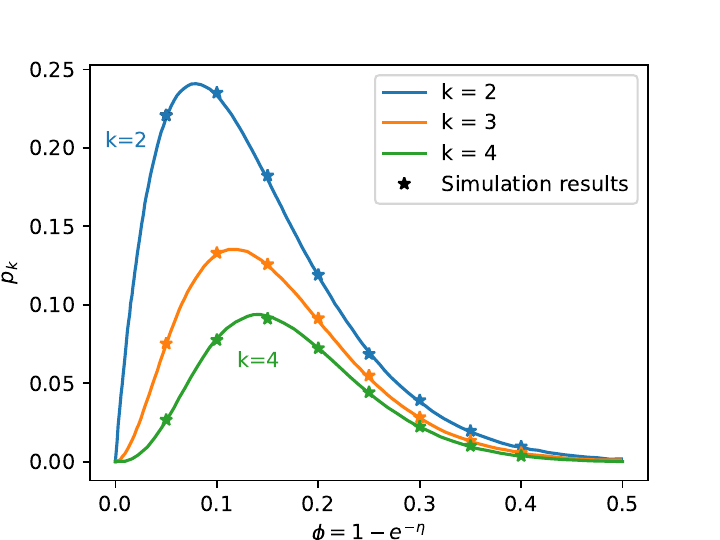}
    \caption{The curves show the theoretical probabilities $p_k$ of a cluster being of size $k$ when picking a random sphere in a system of spheres in 3 dimensions. These are plotted as a function of volume fraction $\phi=1-e^{-\eta}$. At each $\phi$ we have generated $10^5$ clusters using a version of the algorithm where all neighbours of a sphere are visited systemetically, and found the percentage of clusters of size $k=2,3,4$.}
    \label{fig:TestingNeighGen2}
\end{figure}

In order to check that the distribution of cluster sizes matches the theoretical prediction by Ref. \onlinecite{PenroseMathewD1991Oacp,QuintanillaJ1997CiaC}, the algorithm is run until either a cluster of size $1\leq k<5$ has formed or the cluster contains 5 or more spheres. In Fig. \ref{fig:TestingNeighGen2} we plot the results alongside the theoretical predictions.

For a Poisson point process on $\mathbb{R}^d$ with point density $\varrho$, this corresponds to looking at very small volumes of random location until the volume contains the center of a single sphere, i.e. finding a random sphere.

\subsection{The Poisson point process generator algorithm: Constructing the set of points $\mu$ and the assigned set}\label{sec:TheGenAlg}

Let $\kappa_i$ denote the set of assigned points at step $i$, i.e. the set of points that have been assigned neighbours after step $i$, and define the assigned set (or assigned volume), denoted $A_i$, as
$$A_{i} = \bigcup_{\textbf{q}\in \kappa_{i}} \bar B^d_{\textbf{q},2r}.$$ 
Let $\mu_i\subset A_i$ denote the set of points (the centers of spheres) generated by the algorithm after step $i$.

\begin{enumerate}
    \item Set $i:=1$, start at a random point $\mathbf{x}_1\in \mathbb{R}^d$, which can without loss of generality be taken to be $\mathbf{x}_1=0$, and initiate the set of assigned points by letting $\kappa_1 = \{0\}$. Then $A_1=\bar B^d_{0,2r}$. 
    \item Construct a realization $\{y_1,y_2,...,y_k\}$ of a Poisson point process $\sigma_{\textbf{x}_1}$  on $\bar B^d_{0,2r}$,
    where $k$ is a number picked from the Poisson distribution with mean $2^d\eta$. How to do this for $d$--balls will be explained in sec. \ref{sec:ConsSigma}.
    \item Let $\mu_1 = \sigma_{\textbf{x}_1}$. 
    \setcounter{nameOfYourChoice}{\value{enumi}}
\end{enumerate}

\begin{figure}
    \centering
    \includegraphics[width = \linewidth]{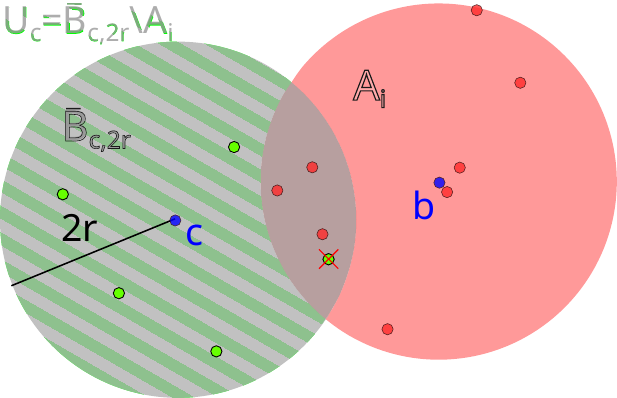}
    \caption{Figure illustrating the concepts of the algorithm and the proof. $\mu_i = \{\text{red dots}\},\ \kappa_i = \{b\},\ \sigma_c = \{\text{green dots}\},\ \sigma_c' = \{\text{green dots excluding those in $A_i$ (i.e. the crossed ones)}\}$. Note that $b$ and $c$ are not points of the Poisson Point process themselves, but only points for which the Poisson Point process is generated around. It is thus not the same as when running the algorithm.}
    \label{fig:Intersecting_volumes}
\end{figure}

\begin{enumerate}
    \setcounter{enumi}{\value{nameOfYourChoice}}
    
    \item Run any chosen dynamic in $\bar B^d_{\textbf{x}_i,r}$ (here a random walk or its inherent dynamics) until a neighbouring sphere is entered. The center of the entered neighbouring sphere will be designated as $\textbf{x}_{i+1}$. Note that in our situation $\textbf{x}_{i+1}\in (\mu_i \cap \bar B^d_{\textbf{x}_i,2r})$. 
    
    \item If $\textbf{x}_{i+1} \in \kappa_i$ then points have already been assigned to the volume $\bar B^d_{\textbf{x}_{i+1},2r}$ defined by $\textbf{x}_{i+1}$. Thus, we set $\mu_{i+1} = \mu_i$ and $\kappa_{i+1} = \kappa_i$, and skip to step 10.
    
    \item Construct a realization of a Poisson point process $\sigma_{\textbf{x}_{i+1}} \subseteq \bar B^d_{\textbf{x}_{i+1},2r}$ as in step 2 (e.g. the green dots in Fig \ref{fig:Intersecting_volumes}).
    
    \item Remove the points of $\sigma_{\textbf{x}_{i+1}}$ that are in $A_{i}$ (e.g. the crossed green dots in Fig. \ref{fig:Intersecting_volumes}), and let $\sigma_{\textbf{x}_{i+1}}'$ denote the collection of remaining points: $$\sigma_{\textbf{x}_{i+1}}' = \sigma_{\textbf{x}_{i+1}}\backslash A_{i}$$
    
    \item Update the assigned volume by adding $\textbf{x}_{i+1}$ to the set of assigned points: $$\kappa_{i+1} = \kappa_i\cup \{\textbf{x}_{i+1}\}$$
    \item Update the set of points by adding the new points $\sigma_{\textbf{x}_{i+1}}'$
    to the set of points in the assigned volume: $$\mu_{i+1} = \mu_i\cup \sigma_{\textbf{x}_{i+1}}'$$

    \item Set $i := i + 1$ and jump to step 4 (or stop the algorithm).
    
\end{enumerate}
When the algorithm stops, we let $A=A_i$ and $\mu=\mu_i$. 
In practice, $A_i$ is not used when running the algorithm, since $\kappa_i$ can be used instead to exclude points in the assigned volume when finding $\sigma_{\textbf{x}_{i+1}}'$. So the only things we really keep track of are the centers of spheres $\mu$ and the assigned points $\kappa.$

\subsection{Definition}
The definition of a Poisson point process can be given to varying degrees of generality, here we give one that is suitable for our purposes, see also 
\cite{Kingman_PoissonProcesses}.

 Let $S \subseteq \mathbb{R}^d$ be a measurable set and let $\Pi$ denote a random, at most countable, collection of points $\Pi\subseteq S$.
 
    Let $\xi(B)$ denote the random variable giving the number of points of $\Pi$ in the measurable set $B\subseteq S$, i.e. $\xi(B)=\#(\Pi\cap B)$.

    Then a Poisson point process on $S$ with density $\lambda > 0$ is a random, at most countable, collection of points $\Pi\subseteq S$ with the following properties:
    \begin{itemize}
        \item For every measurable $B\subseteq S$, the probability that 
        $\xi(B)=k$ is given by 
        \begin{equation}
            P(\xi(B)=k)=\frac{(\lambda |B|))^k}{k!}e^{-\lambda |B| }
            \label{eq:PPPprop}
        \end{equation}
        where $|B|$ is the Lebesgue measure ($d$-"volume") of $B$. In other words, $\xi(B)$ is Poisson distributed with mean $\lambda |B|$. \footnote{    The Poisson distribution property derives from the fact that the existence of a point on each site in $B$ is binomial with probability $p$ and that the Poisson distribution is the limit of the binomial distribution for constant $Np=|B|\lambda$ when $N\to \infty$ and $p=|B|\lambda/N\to 0$, where N is the number of sites in $B$.}
        \item For any disjoint measurable subsets $B_1$, $B_2$ of $S$, the random variables $\xi(B_1)$ and $\xi(B_2)$ are independent.
    \end{itemize}

Since the density $\lambda$ is constant, this is an example of a homogeneous Poisson point process. In a more general definition of a Poisson point process, the density is allowed to depend on the position in space, but this is not relevant for our application.

We need some important (derived) properties of the Poisson point process, which in our context can be formulated as follows:

\subsubsection{Restriction Property}
Let $\Pi$ be a Poisson point process with density $\lambda>0$ on $S$ and let $S'$ be a measurable subset of $S$. Then $\Pi'=\Pi\cap S'$ is a Poisson point process on $S'$ with density $\lambda$. This property follows immediately from the definition of a Poisson point process (see also Restriction Theorem \cite[Sect. 2.2]{Kingman_PoissonProcesses}).

\subsubsection{Superposition property}
Let $S_1, S_2\subseteq \mathbb{R}^d$ be disjoint measurable sets and let $\Pi_1,\Pi_2$ be independent Poisson point processes, both with density $\lambda>0$, on $S_1$, $S_2$ respectively. Then the union $\Pi=\Pi_1\cup \Pi_2$ is a Poisson point process on $S_1\cup S_2$ with density $\lambda$. This property follows from the additivity property of Poisson distributions and is a special case of the much more general Superposition theorem (see again \cite[Sect. 2.2]{Kingman_PoissonProcesses}).

\subsubsection{Conditional distributions}\label{secconditional}
The distribution of neighbours for an arbitrary point of a Poisson point process and the size of a cluster containing an arbitrary sphere in the hypersphere model, are both described using conditional distributions in the sense of Palm distributions. 

Thus, a Poisson point process with density $\lambda$, has the property that the number of "neighbours" of an arbitrary point, in a measurable $B\subset S$ containing the point, is Poisson distributed with mean $\lambda|B|$. This is given by the so-called reduced Palm distribution, cf. \cite{Stoyan1}. In particular, the number of neighbours (that is, sphere centers) of an arbitrary sphere in the hypersphere model, in a sphere of radius $2r$, is Poisson distributed with mean $2^d\eta$. 

More generally, the conditional distribution of a homogeneous Poisson process, given that there is a point at a certain location, which can without loss of generality be assumed to be 0, is called the Palm distribution. This conditional distribution is the same as the distribution of the original process, together with a point at 0. Adding a point at 0 thus  corresponds to considering an arbitrary point of the process, which is without loss of generality assumed to be at 0. See f.ex. \cite{Stoyan1} and \cite{PenroseMathewD1991Oacp}.

\subsection{Proof}\label{secproof}
\noindent We want to show that the algorithm produces a realization of a Poisson point process $\mu$ on the set $A$ with density $\varrho$. We do this by induction, showing that for every $i\in\mathbb{N}$, $\mu_i$ is a Poisson point process on $A_i$ with density $\varrho$ (where $\mu_i=\Pi$, $A_i=S$ and $\varrho=\lambda$ in the notation of the definition). 
   
    By construction $\mu_1=\sigma_{\textbf{x}_1}$ is a Poisson point process on $A_1=B^d_{\textbf{x}_1,2r}$ with density $\varrho$, so the statement is true for $i=1$.
    
    We now show that if $\mu_i$ is a Poisson point process on $A_i$, then $\mu_{i+1}$ is a Poisson point process on $A_{i+1}$:
    \begin{enumerate}
        \item  By construction $\sigma_{i+1}$ is a Poisson point process on $B^d_{\textbf{x}_{i+1},2r}$ with density $\varrho$.
        
        \item Define $U_{\textbf{x}_{i+1}} = B^d_{\textbf{x}_{i+1},2r}\backslash A_{i}$. This is the "new volume" to which points have been assigned in step $i+1$.
        
        \item By the restriction property $\sigma_{\textbf{x}_{i+1}}' =\sigma_{\textbf{x}_{i+1}} \bigcap U_{\textbf{x}_{i+1}}$ is a Poisson point process on $U_{\textbf{x}_{i+1}}$ with density $\varrho$, with $U_{\textbf{x}_{i+1}}$ in the role of $S'$ and $\sigma_{\textbf{x}_{i+1}}'$ in the role of $\Pi'$. 
        \item Recall that $A_{i+1} = A_i \cup B^d_{\textbf{x}_{i+1,2r}}$ so that $A_{i+1}$ is the union of the disjoint sets $A_i$ and $U_{\textbf{x}_{i+1}}$: $A_{i+1} = A_i \cup U_{\textbf{x}_{i+1}}$.
        
        
        \item Since $\mu_i$ and $\sigma_{\textbf{x}_{i+1}}'$ are independent Poisson point processes on $A_i$ and $U_{\textbf{x}_{i+1}}$, respectively, both with density $\varrho$, it follows from the Superposition property that 
$\mu_{i+1} = \mu_i\cup \sigma_{\textbf{x}_{i+1}}'$ is a Poisson point process on $A_{i+1}$ with density $\varrho$. (Here with $\mu_i$ and $\sigma_{\textbf{x}_{i+1}}'$ in the roles of $\Pi_1$ and $\Pi_2$ and $A_i$ and $U_{\textbf{x}_{i+1}}$ in the roles of $S_1$ and $S_2$).

    \end{enumerate}   

Thus the algorithm produces a set $A$ and a realization of a Poisson point process $\mu$ on $A$, with its system of spheres. 

Adding the point $\textbf{x}_1$ at 0 produces the conditioned Poisson process, given there is a point at 0.  This corresponds to taking as starting point for the algorithm an arbitrary point of the Poisson point process, which we can assume without loss of generality to lie at 0.

In particular, the algorithm produces a cluster of spheres containing the sphere at 0, of a given size. Each realization occurs with the same probability as in the Poisson "blob" model, because of the independence property of Poisson point processes. This is illustrated in Figure \ref{fig:TestingNeighGen2}, which shows the probabilities that the cluster containing 0 produced by the algorithm is of size 2, 3 and 4, together with theoretical probabilites from the Poisson blob model, in 3 dimensions.

\subsection{Constructing $\sigma_\mathbf{x}$}
\label{sec:ConsSigma}
The aim of this section is to construct the Poisson point process $\sigma_\mathbf{x}$ on $B^n_{\textbf{x}_1,2R}$ with point density $\varrho$. Note that in the section we use $n$ instead of $d$ to refer to the number of dimensions in order to avoid confusion with differentials. In order to construct $\sigma_\mathbf{x}$, we first find a point on the boundary of an n ball of radius \textit{ r}. We then find the distance from the center of a sphere of radius R to any random point inside the sphere. Using the found random point on the boundary of a sphere of the radius given by the found distance then yields the location of a single random point inside the sphere of radius R. Using the definition of a Poisson point process, we then construct $\sigma_\mathbf{x}$.

\subsubsection{\label{sec:randpoint2}Finding a random point inside the volume of an $n$-ball}
Let $r$ denote the radius of an $n$-ball centered at $\mathbf{x}$. Let $X_1,X_2,...,X_n$ be $n$ independent normal random variables with mean 0 and variance 1. A random point on the boundary of $B_{\mathbf{x},r}^n$ is a random element $Y_r^\mathbf{x}$ given by\cite{MullerMervin1959Anoa}, 
\begin{equation}
\label{eq:randpointonsphere}
    Y^\mathbf{x}_r=\mathbf{x}+\frac{r}{\sqrt{X_1^2+X_2^2+...+X_n^2}}\left(X_1, X_2,...,X_n\right).
\end{equation}

We thus only need to find a random variable describing the distance from the center of the n-ball to a random point inside of it. The probability $p_1(r)$ of finding a point at distance $r$ from the center of a hypersphere of radius R is given by:
\begin{equation}\label{eq:PPPROPDX}
    p_1(r) = \frac{S_n(r)}{V_n(R)}, \quad r\in[0,R]
\end{equation}
where $S_n(r)$ is the surface area of a $n$-ball of radius $r$ and $V_n(R)$ the volume of a hypersphere of radius R. These are given by
\begin{equation}
    S_n(r) = \frac{2\pi^{n/2}}{\Gamma(\frac{n}{2})}r^{n-1}\quad \wedge \quad V_n(R) = \frac{\pi^{n/2}}{\Gamma(\frac{n}{2}+1)}R^n,
\end{equation}
with $\Gamma$ being the gamma function. 
Inserting these into equation (\ref{eq:PPPROPDX}) and integrating yields the cumulative distribution function for $p_1(r)$:
\begin{align}
    P_1(Z\leq r)&=\int_0^r p_1(s)ds \\
    &= \int_0^r \frac{2\Gamma(\frac{2}{n}+1)}{\Gamma(\frac{2}{n})}\frac{s^{n-1}}{R^n}ds\\
    &=\int_0^r n\frac{s^{n-1}}{R^n}ds = \frac{r^n}{R^n}
\end{align}
Letting $S$ be uniformly distributed in [0,1] (that is $p_2(S) = 1$ for any $S\in[0,1]$) yields the cumulative distribution function:
\begin{equation}
    P_2(S\leq y) = \int_0^y p_2(s)ds = y
\end{equation}
We thus have that  $y = r^n/R^n$ when the cumulative distribution functions are equal (i.e. $P_1=P_2=P$). Therefore,
\begin{equation}
    P\left(S\leq \frac{r^n}{R^n}\right) = P(RS^{1/n}\leq r) = P(Z\leq r)
\end{equation}
Thus, if S is a random variable uniformly distributed in [0,1], then 
\begin{equation}\label{eq:randradius}
    Z = RS^{1/n}
\end{equation}
is a random variable describing the distance from the center of the $n$--ball to a random point inside of it. Combining Eq. (\ref{eq:randpointonsphere}) and (\ref{eq:randradius}) yields the following result:

Let $S$ be a uniform random variable in $[0,1]$, and $X_1,X_2,...,X_n$ be $n$ independent normal random variables with mean 0 and variance 1. A random point $X_R^\mathbf{x}$ in $\bar{B}_{\mathbf{x},R}^n$ is given by:
\begin{equation}
\label{eq:randpoint}
        X^\mathbf{x}_R=\mathbf{x}+\frac{RS^{1/n}}{\sqrt{X_1^2+X_2^2+...+X_n^2}}\left(X_1, X_2,...,X_n\right)
\end{equation}

\subsubsection{Constructing a realization of $\sigma_x$}
\label{sec:ConsSigma2}

According to the definition of the Poisson point process, the number of points in $\sigma_\mathbf{x}$ is naturally the Poisson random variable $\xi(B_{\mathbf{x},2R}^n)$ with mean
\begin{align}
    \label{eq:meanNeighbours}
    \varrho |B_{\mathbf{x},2R}^n|=\varrho V_n(2R)=\varrho \frac{\pi^\frac{n}{2}}{\Gamma(\frac{n}{2}+1)}(2R)^n =2^n\varrho V_n(R) = 2^n\eta\,.
\end{align}
Here the first three equalities follow from the volume of an $n$-ball and the last equality follows from the definition of reduced density.

Let $y_1,...,y_k$ be $k=\xi(B_{\mathbf{x},2R}^n)$ random points on $B_{\mathbf{x},2R}^n$, each given by Eq. (\ref{eq:randpoint}). Since $y_1,...,y_k$ are independently distributed on $B_{\mathbf{x},2R}^n$
\begin{equation}
    \label{eq:sigma}
    \{y_1,...,y_k\}
\end{equation}
is a realization of a Poisson point process $\sigma_\mathbf{x}$ on $\bar{B}_{\mathbf{x},2R}^n$.

\section{\label{appendix:Rates}Transition rates.}

To apply the kinetic Monte Carlo algorithm, an expression for $\Gamma_{ij}$, the transition rate between two intersecting hyperspheres, $i$ and $j$ is needed. To find this rate,  random walks on a system consisting of two intersecting unit hyperspheres were performed for $d=12$ and $d=18$ respectively. The distance between the two spheres is denoted $L_{ij}$ and the radius of the intersection is denoted $h_{ij}$, see Fig. 4 of the main text.

\begin{figure}
    \centering
    \includegraphics[width=\linewidth]{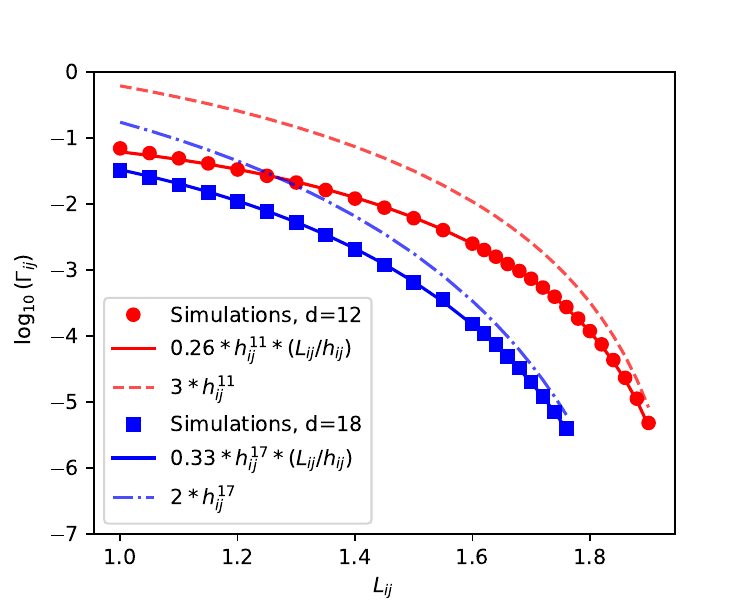}
    \caption{
    Transition rates between two unit spheres with centers separated by $L_{ij}$ in 12 (red) and 18 (blue) dimensions. Points: simulation results from random walks. Full lines: Fits  to Eq. (\ref{eq:jump_probability}) of the main text. Dashed lines: $\Gamma_{ij} \propto h_{ij}^{d-1}$ shown for comparison.
    \label{fig:TransitionRates}
    }
\end{figure}

17000 random walkers were started in sphere $i$ (excluding the part that is closest to the center of sphere $j$). Steps in the random walk was chosen with equal probability on the surface of a d-dimensional hypersphere of radius 0.01. 
The fraction of walkers being in sphere $i$ was monitored as a function of time. As expected, an exponential decay from fraction 1 to 0.5 was found. The transition rates determined from the exponential decays are plotted as points in Fig.~\ref{fig:TransitionRates}. 

Since the random walk has equal probability per volume, the rates are expected to be proportional to the volume of the intersection relative to the volume of the initial sphere:
\begin{equation}
    \Gamma_{ij} \propto h_{ij}^{d-1} dL
\end{equation}
where $dL$ is the effective width of the intersection. Setting $dL$ proportional to the inverse slope of the hypersphere surface at the intersection (see Fig. 4 of the main text), $dL \propto L_{ij}/h_{ij}$, leads to Eq. (\ref{eq:jump_probability}) of the main text, which is found to fit the simulation results very well; see Fig.~\ref{fig:TransitionRates}.

\newpage

\providecommand{\noopsort}[1]{}\providecommand{\singleletter}[1]{#1}%

\end{document}